\documentclass[letterpaper, 10 pt, conference]{ieeeconf}
\IEEEoverridecommandlockouts

\usepackage{url}
\usepackage{tcolorbox}
\usepackage{xcolor}
\usepackage{cite}
\usepackage{amsmath,amssymb,amsfonts}
\usepackage{algorithmic}
\usepackage{graphicx}
\usepackage{textcomp}
\usepackage{xcolor}
\usepackage{lipsum}
\usepackage{booktabs}
\usepackage{placeins}
\usepackage{afterpage}
\usepackage{colortbl}%
  
\usepackage[caption=false,font=normalsize,labelfont=sf,textfont=sf]{subfig}

\FloatBarrier
\definecolor{mygreen}{RGB}{0, 0, 0}
\def\BibTeX{{\rm B\kern-.05em{\sc i\kern-.025em b}\kern-.08em
    T\kern-.1667em\lower.7ex\hbox{E}\kern-.125emX}}
    
\begin{document}

\title{\LARGE \bf
Conditional Electrocardiogram Generation \\ Using Hierarchical Variational Autoencoders
}

\author{Ivan Sviridov$^{1}$, Konstantin Egorov$^{1}$
\thanks{$^{1}$Sber AI Lab, Moscow, Russia}
\thanks{{email: \textcolor{blue}{wchhiaarid@gmail.com}}}}

\maketitle

\begin{abstract}
Cardiovascular diseases (CVDs) are disorders impacting the heart and circulatory system. These disorders are the foremost and continuously escalating cause of mortality worldwide. One of the main tasks when working with CVDs is analyzing and identifying pathologies on a 12-lead electrocardiogram (ECG) with a standard 10-second duration. Using machine learning (ML) in automatic ECG analysis increases CVD diagnostics' availability, speed, and accuracy. However, the most significant difficulty in developing ML models is obtaining a sufficient training dataset. Due to the limitations of medical data usage, such as expensiveness, errors, the ambiguity of labels, imbalance of classes, and privacy issues, utilizing synthetic samples depending on specific pathologies bypasses these restrictions and improves algorithm quality. Existing solutions for the conditional generation of ECG signals are mainly built on Generative Adversarial Networks (GANs), and only a few papers consider the architectures based on Variational Autoencoders (VAEs), showing comparable results in recent works. This paper proposes the publicly available conditional Nouveau VAE model for ECG signal generation (cNVAE-ECG), which produces high-resolution ECGs with multiple pathologies. We provide an extensive comparison of the proposed model on various practical downstream tasks, including transfer learning scenarios showing an area under the receiver operating characteristic (AUROC) increase up to $2\%$ surpassing GAN-like competitors.
\textcolor{blue}{\url{https://github.com/univanxx/cNVAE_ECG}}
\newline

\end{abstract}

\section{Introduction}\label{intro}
Cardiovascular diseases (CVDs) are a family of pathologies that affect the heart and blood vessels. These diseases are the leading and annually growing cause of death worldwide~\cite{pmid36695182}. Given their prevalence and impact, ongoing research is required to identify causes and risk factors, develop effective treatment and prevention strategies, and create technologies to find and monitor heart diseases. Electrocardiogram analysis is one of the leading areas of research that can detect and prevent CVDs. 

Electrocardiography (ECG) is a non-invasive diagnosis and heart activity monitoring method. The main goal of this method is to record the processes of electrical depolarization and repolarization and generate a graphical representation of this activity in the form of a signal. Each signal represents the potential difference between electrodes attached to different human body parts. This difference is called a lead. Acquired information allows for analyzing the work of the heart, identifying deviations, and determining the frequency and rhythm of heart contractions. 

Deep learning is widely used to analyze and process ECG signals, as it allows for the automatic detection of abnormalities and pathologies~\cite{LIU2021107187}. This approach allows for high-quality diagnosis of heart diseases and monitoring of the patient’s condition. Also, deep learning algorithms make it possible to automate the process of interpretation and detection of pathologies. However, a large amount of high-quality data is crucial for the correct work of deep learning algorithms. But in medical machine learning tasks, in particular, when working with ECG signals, some specific problems arise when obtaining high-quality data, namely:

\begin{itemize}
\item \textbf{Limited access to data}: In medical research, access to large volumes of data may be restricted due to issues with patient confidentiality, ethical considerations, and legislation \cite{GERKE2020295}. These peculiarities can significantly limit the amount of data available for training algorithms. 
\item \textbf{Imbalanced data}: Medical data often has a strong class imbalance. For example, even in large ECG datasets~\cite{9662687}, some classes can be heavily underrepresented with only a pair of instances, which makes it challenging to train algorithms with high generalization ability.
\item \textbf{Poor quality of data}: Data quality can be an issue in medical applications. Errors in data collection, noise, missing values, or inaccurate labels can affect the performance of algorithms and lead to incorrect conclusions and actions. When handling ECG signals, the common causes of errors include body movements during measurement procedure, poor electrode contact, skin-electrode impedance, etc.~\cite{43620}.
\item \textbf{Complexity of data creation}: All medical data results come from the collaboration between the patient, the medical specialist, and the medical device. It is necessary to carry out a corresponding measurement of a person with special equipment with the help of a professional to obtain at least one data sample. This problem makes creating a dataset a time-consuming and investment-intensive process.
\end{itemize}

Generating synthetic data can solve the problems described above. The output can be obtained in a required amount of high-quality data for a specific task if an approach to the training of the ECG signal synthesis algorithm is used correctly. This process makes it possible to lower class imbalance by generating data for all classes and reducing the cost of obtaining new samples. In particular, to get a high-quality result of ECG signal classification, it is necessary to take into account several aspects, namely, 1) generate all main 12 leads with a standard length of 10 seconds~\cite{doi:10.1161/CIR.0000000000000025} and 2) use conditional generation based on specific classes to obtain results for different pathologies.

The most common approach for generating ECG signals is to use the models based on  Generative Adversarial Networks (GANs), although modern architectures based on Variational Autoencoders (VAEs)~\cite{kingma2022autoencoding} such as Nouveau VAE (NVAE) model~\cite{vahdat2021nvae} also show comparable results for image generation tasks. However, there is no work on conditional ECG generation by VAE-based models and verification of the quality of generated samples on practical downstream tasks. To close this research gap, in this work, we propose the novel model named conditional NVAE for ECG or cNVAE-ECG based on NVAE for conditional ECG generation with publicly available code and show that using generated data by cNVAE-ECG demonstrates the best test metrics in the two downstream tasks of identifying pathologies and transfer learning on ECG signals. We also compare the developed algorithm with state-of-the-art GAN-based methods. In particular, for a binary classification problem, adding cNVAE-ECG generated signals to a train set gives an increase of up to 2\% in the area under the receiver operating characteristic (AUROC) metric, outperforming GAN-like approaches, and for a multi-label classification problem, using a pre-train enriched with cNVAE-ECG generated signals in most cases gives best result according to AUROC.

The structure of our paper is as follows: Section~\ref{rel_work} describes existing approaches for generating ECG signals and their features. Section~\ref{nvae_mdl} introduces the cNVAE-ECG model, based on NVAE for conditional generation of ECG signals. Section~\ref{pipeline} describes an Experimental Setup that tests the improvement in the metrics of two downstream tasks when adding generated data and comparing it with GAN-based methods. Section~\ref{results} demonstrates the results of the described experiments, and conclusions about the results obtained and further plans are described in Section~\ref{conclusion}.

\section{Related Work}\label{rel_work}
\subsection{ECG Generation}

The most common methods for generating ECG signals are models based on GANs. For example,~\cite{Golany_Radinsky_2019} solved the task of generating one lead and single heartbeat ECG signals and proposed an architecture named Personalized Generative Adversarial Network (PGAN). The discriminator of this model contains two cross-entropies to maximize the log probability of assigning the correct labels to both classes, and the generator is additionally penalized by MSE loss according to the ECG signal structure. Another work~\cite {9389779} proposed the ProEGAN-MS architecture, a GAN model divided into four stages, each of which is a more complex version of the previous stage. Also, \cite{pmlr-v119-golany20a} used a system of Ordinary Differential Equations (ODE) representing heart dynamics and incorporated this ODE system into the optimization process of a standard GAN network to create biologically correct ECG training examples.

However, generating only one class can be a bottleneck since, to improve classification quality, it is essential to generally increase the training set size so that the model can capture more patterns and dive deeper into the data. Also, creating a separate model to generate each class may miss dependencies between them and become a memory-intensive process. Thus, \cite{9207613} used GAN to conditionally generate one lead for several signal types, adding a class label to the generator and discriminator so that the model understood class structure.

The next bottleneck in working with ECG signals is the duration and number of leads. In previous articles, the authors worked with just one heartbeat and one lead, which allows the model to capture relationships in the data less accurately than when working with multiple leads and a more extended sequence. Therefore, \cite{Thambawita2021-mw} proposed the GAN-like models named WaveGAN* and Pulse2Pulse models to generate the main 12 leads with a length of 5000 (using signals with the sampling frequency of $500$ Hz and $10$ seconds duration). This approach made it possible to study the depth of interaction between generative models and multi-channel sequences. 

Despite the abovementioned approaches, only one paper avoids all bottlenecks. It describes the conditional generation of ECG signals to check that the test metrics on downstream tasks increase when enriching the training set with generated signals. Thus, \cite{10095035} developed the MLCGAN model based on WaveGAN enriched with convolutions, which showed that creating such a GAN-like model for the mentioned task is possible. 

Although there is much work on using GAN-like models, there are papers on generating ECGs for other architectures. For example, in the work~\cite{alcaraz2023diffusionbased}, the authors proposed the architecture based on diffusion models~\cite{DBLP:journals/corr/abs-2006-11239}. They showed that this architecture is capable of conditionally generating ECG signals but without testing on the downstream task. However, this model type requires significant training costs, a substantial limitation for practical use~\cite{dhariwal2021diffusion}.

There are also several works on ECG signal generation using another type of architecture named VAE. For example, \cite{XIA2023104587} compared the quality of conditional generation of 12 leads ECG signals with 1-second duration in GAN-like and VAE-like architectures and showed that the conditional generative framework based on VAE can shorten the training time and simplify the generation process without significant performance loss and demonstrates similar potential to the GAN-based models. Also, research on the field of comparison VAEs and GANs in the generation of ECG signals \cite{10.1007/978-3-030-22885-9_1} showed that variational autoencoders are easy and fast to train and infer compared to GAN-like models, but they usually perform worse than them.

There are also works devoted to generating ECG signals using VAE-like models. Thus, \cite{nn_ecg} proposed a VAE-convolutional model to generate 12 leads of 10-second ECGs utilizing a set of features. In the work \cite{9761431}, authors also conditionally generated signals of the same structure, using VAE with convolutions, and showed that the generation results are of high quality and can be used in further work. \cite{9761590} proposed a novel multimodal VAE capable of processing combined physiology and anatomy information in the form of one lead of one beat ECGs and 3D point clouds, which showed the possibility of VAE to work with multiple data dimensions and producing quality results.

Even though models based on VAEs show results similar to models based on GANs, there is limited work on the conditional generation of ECG signals by VAE-like architectures. Based on the assumption that improving the quality of these models is possible, our work aims to develop a model named cNVAE-ECG based on an existing VAE-like architecture for conditional ECG generation, namely the NVAE model, to solve problems of improving classification quality. Also, we test and compare cNVAE-ECG with existing GAN-based models on the downstream tasks of pathologies identification and transfer learning.


\subsection{NVAE Background}

\subsubsection{Variational Autoencoders}

Variational Autoencoders are stochastic architectures that use variational inference. These types of models are capable of approximating a data space $\bold{x}$ having a distribution $p_{\theta}(\bold{x})$ parameterized by $\theta$ and generating new samples from that space using hidden (latent) representation $p_{\theta}(\bold{z})$.

An autoencoder is a neural network consisting of two blocks: an encoder and a decoder. At training, an encoder reduces distribution $p_{\theta}(\bold{x})$ to compressed (latent) representation $p_{\theta}(\bold{z} \vert \bold{x})$ using nonlinear transformations. After sampling from this representation, the decoder tries to construct the original representation $p_{\theta}(\bold{x} \vert \bold{z})$. At inference, only the decoder reconstructs $p_{\theta}(\bold{x} \vert \bold{z})$ after sampling from $p_{\theta}(\bold{z})$.

The disadvantage of lower-quality generation compared to GANs mentioned earlier can be reduced using advanced techniques in variational inference, which can generate higher-quality sample data simultaneously, and the decision is to use deep hierarchical VAE architecture. In hierarchical VAE, latent variables are partitioned into disjoint groups $\textbf{z}=\{z_1, z_2, ...,z_L\}$, where $L$ is the number of groups. Hence, prior is represented by $p(\textbf{z})=\prod_{l}p(\textbf{z}_l | \textbf{z}_{l-1} )$ and the approximate posterior $q(\textbf{z}\|\textbf{x})=\prod_{l}(\textbf{z}_l | \textbf{z}_{l-1}, \textbf{x})$, where factorial Normal distributions represent each conditional prior and posterior.

The main problem of this approach is the instability of architecture. In such approaches, many collapses are possible, such as the attenuation of higher layers of the hierarchy or gradient explosions. To solve these problems, including the problem of the quality of generation of variational autoencoders, the NVAE model was designed.

\subsubsection{NVAE Model}

NVAE is a deep hierarchical VAE initially built for image generation. This architecture looks like a hierarchy, where the encoder is a bottom-up model, which transforms data into spatial representation, and the decoder (or generative block) is a top-down model, which goes from high-dimensional features to original data. With new hierarchy levels, the model goes from short-range to long-range correlations. The authors of NVAE improved the existing architecture with some techniques, making a stable model with relatively fast training and inference. Some of these techniques are described below.

\textbf{Mixture of discretized Logistic distribution}: In the NVAE, authors used the approach described in the PixelCNN++ by \cite{salimans2017pixelcnn}, which is that each pixel of the output image from $p(\textbf{x}|\textbf{z})$ is predicted through a mixture of discretized Logistic distributions. This technique speeds up learning and inference and gives a smoother output image.

\textbf{Channel Dependence}: The authors also used the PixelCNN++ approach to generate output so that the values of the three RGB channels depend on each other. That is, the very first channel, red, which is generated by a mixture of Logistic distributions, has parameters $c_R$ independent of the others, and the green channel is specified by a linear combination of its mixture of distributions and parameters of the red channel $c_G + \alpha \cdot c_R$ . The blue channel is a linear combination of the red and green channels with its mixture generation results: $c_B + \beta \cdot c_G + \gamma \cdot c_R$. This generation approach strengthens the connections between channels and allows the generation of higher-quality samples.

NVAE model contains these and other improvements in hierarchical VAE, which they tested on multiple datasets, such as MNIST \cite{6296535}, CIFAR-10 \cite{Krizhevsky2009LearningML}, CelebA 64 \cite{larsen2016autoencoding}, and CelebA HQ \cite{karras2018progressive}, obtaining state-of-the-art results among non-autoregressive likelihood-based models. However, the described datasets represent a collection of one- and three-dimensional channels. This paper describes the cNVAE-ECG, an NVAE-based architecture that can conditionally generate ECG signals.

\section{Proposed Approach} \label{nvae_mdl}

This section proposes a cNVAE-ECG - NVAE-based model for working with ECG signals. To conditionally generate standard ECG, it was necessary to develop an NVAE-based architecture capable of working with 12-channel 1D signals and generating according to a given class.

 \subsection{ECG generation}

As mentioned in Section~\ref{intro}, standard ECG has 12 main leads - namely six limb leads (I, II, III, aVR, aVL, aVF) and six chest leads $(\text{V}_1, \dots, \text{V}_6)$. One of the main properties of such a representation is that the leads of the limbs are, by their nature, interconnected by the relationships \cite{inbook} called Einthoven's law:

\begin{equation}
    \text{I}+\text{III}=\text{II}
\label{eq:Einthoven}
\end{equation}

and Goldberger's equations: 

\begin{equation}
  \begin{split}
    aVL = \frac{\text{I}-\text{III}}{2} \\ 
    -aVR = \frac{\text{I}+\text{II}}{2} \\
    aVF = \frac{\text{III}+\text{III}}{2}
  \end{split}
\label{eq:Goldberger}
\end{equation}

Therefore, using laws \ref{eq:Einthoven} and \ref{eq:Goldberger}, the generation of six limb leads I, II, III, aVR, aVL, and aVF can be replaced by the generation of only two leads, I and III. Thus, we obtained 8 leads during the work of the model and then transformed them into 12 main leads. To generate these leads, we determined the procedure for working with mixtures and generating channels, which was mentioned in Section~\ref{rel_work}. Initially, each of the Logistic distributions $P_L(m_1),..,P_L(m_K)$ in a mixture returns outputs using inverse sampling:

\begin{equation}
    P_L(m_i) = \mu_{m_i}+s_{m_i} \, \text{}{log}\left( \frac{u}{1-u} \right),
\label{eq:logistic}
\end{equation}

where $\mu_{m_i}$ and $s_{m_i}$ are parameters of $m_i^{\text{th}}$ Logistic distribution, $u \sim \text{Uniform(0,1)}$.

For this purpose, we proposed the algorithm for constructing mixtures distribution to work with eight one-dimensional ECG signals, or more precisely, we redefined the method for selecting distribution parameters. The relationship between channels or leads was determined as follows using Eq.~\ref{eq:logistic}:

\begin{enumerate}
\item The first limb lead I is generated as $P_L(\text{I}_i|C_{\text{I}_i})=P_L(\text{I}_i|\mu_{\text{I}_i}(C_{\text{I}_i}), s_{\text{I}_i}(C_{\text{I}_i}))$, where $C_{\text{I}_i}$ is the context tensor obtained by Logistic mixture for the lead I. 
\item After this, limb lead III is generated with parameters $\mu_{\text{III}_i}(C_{\text{III}_i}, \text{I}_i)=\mu_{\text{III}_i}(C_{\text{III}_i})+\beta\mu_{\text{I}_i}(C_{\text{I}_i})$ and $s_{\text{III}_i}(C_{\text{III}_i}, \text{I}_i)=s_{\text{III}_i}(C_{\text{III}_i})+\beta s_{\text{I}_i}(C_{\text{I}_i})$.
\item The chest lead $\text{V}_1$ is then generated independently of the previous limb leads with the parameters $\mu_{\text{V}_{1,i}}(C_{\text{V}_{1i}})$ and $s_{\text{V}_{1i}}(C_{\text{V}_{1i}})$ since the chest leads provide information about vertical planes, and the limb leads to focus on horizontal planes.

\item Then, Lead $\text{V}_2$ is generated with parameters $\mu_{\text{V}_{2,i}}(C_{\text{V}_{2,i}},V_{1,i})=\mu_{\text{V}_{2,i}}(C_{\text{V}_{2,i}})+\alpha({\text{V}_{2,i}}) \cdot \mu_{\text{V}_{1,i}}(C_{\text{V}_{1,i}})$, $s_{\text{V}_{2,i}}(C_{\text{V}_{2,i}},V_{1,i})=s_{\text{V}_{2,i}}(C_{\text{V}_{2,i}})+\alpha({\text{V}_{2,i}}) \cdot s_{\text{V}_{1,i}}(C_{\text{V}_{1,i}})$.
\item The remaining leads are generated according to the formula $\\ \mu_{\text{V}_{k,i}}(C_{\text{V}_{k,i}}, V_{k,i},V_{k-1,i}, \cdots ,V_{1,i})=\mu_{\text{V}_{k,i}}(C_{\text{V}_{k,i}})+\sum_{j=1}^{k-1} \alpha({\text{V}_{j_k,i}}) \cdot \,\mu_{\text{V}_{j,i}}(C_{\text{V}_{j,i}})$, $s_{\text{V}_{k,i}}(C_{\text{V}_{k,i}}, V_{k,i},V_{k-1,i}, \cdots ,V_{1,i})=s_{\text{V}_{k,i}}(C_{\text{V}_{k,i}})+\sum_{j=1}^{k-1} \alpha({\text{V}_{j_k,i}}) \cdot \,s_{\text{V}_{j,i}}(C_{\text{V}_{j,i}})$.
\end{enumerate}

Another property of ECG is that the representation of each signal is one-dimensional and has a length of 5000 (sampling frequency of $500$ Hz and $10$ seconds duration). Therefore, we decided to implement one-dimensional convolutions to improve the quality of the generated signals since this makes it possible to work with signals directly without using transformations to a two-dimensional form, as in some of the works from Section~\ref{rel_work}. However, since the model was initially designed to work with two-dimensional tensors, we first converted the ECG signals to a two-dimensional representation using a short-time Fourier transform or STFT \cite{1164317}. As a result, each signal was a two-channel and two-dimensional spectrogram, the first channel of which was the real part of the Fourier transform, and the second was the imaginary part. In Section~\ref{results}, we demonstrate that this approach called 2D-cNVAE-ECG showed that the metric on the downstream task does not improve when adding samples generated in this approach but even decreases. Also, we modified several hyperparameters, such as the kernel size, and additionally introduced operations for dimension alignment in our NVAE implementation to enable the processing of signals with a length of 5000.

 \subsection{Conditional generation}

The essence of the conditional generation problem is that a single model is used to obtain a sample of a specific class rather than building several independent architectures. This approach allows the model to learn global dependencies in the data and evaluate each class feature without losing the dependency between them. One of the most common approaches is to add a class representation in the form of embedding to the architecture, for example, as in works \cite{9207613,9761431,alcaraz2023diffusionbased,10095035} with a conditional generation of ECG signals.

\begin{figure}[ht]
\begin{center}
\centerline{\includegraphics[width=\columnwidth]{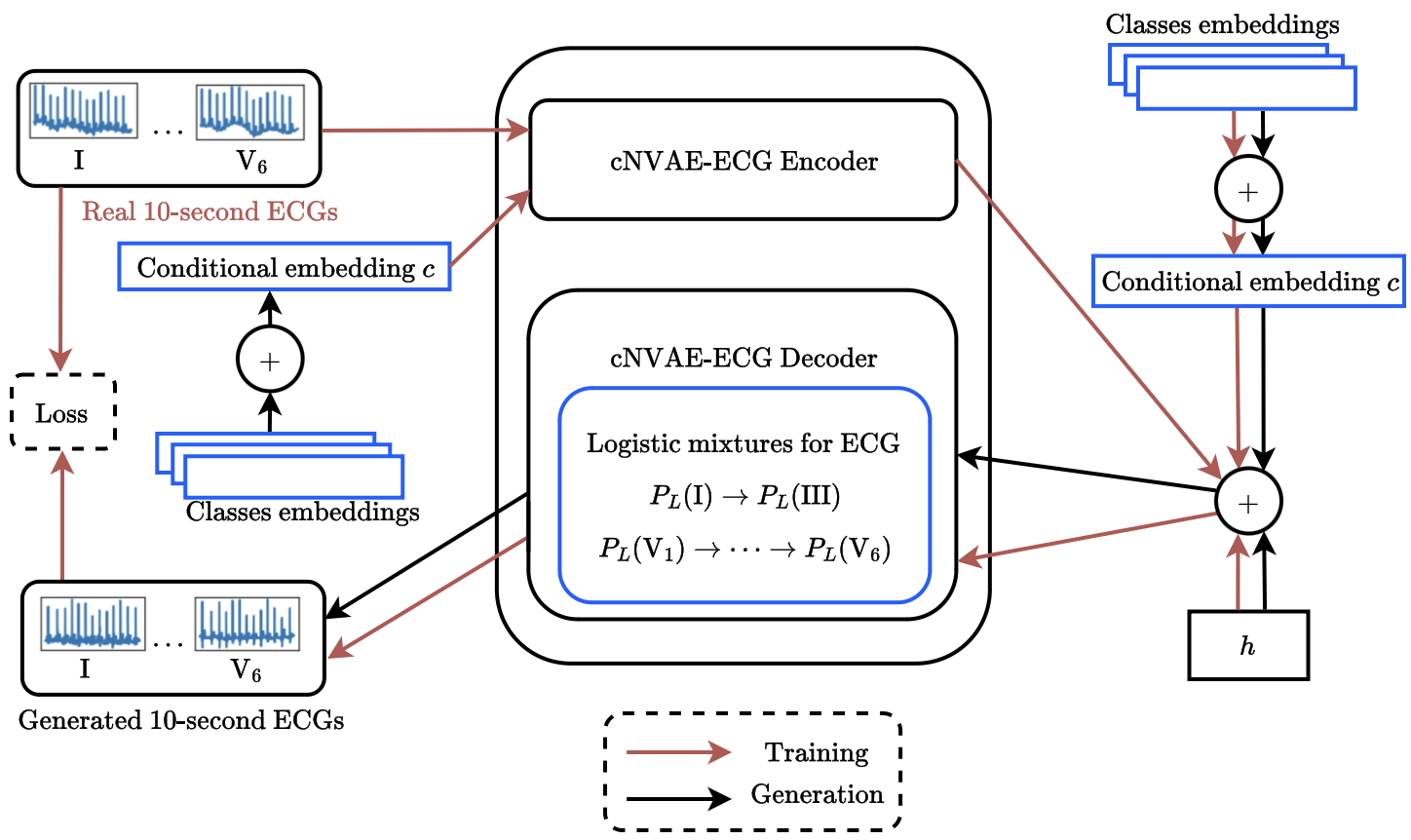}}
\caption{Proposed cNVAE-ECG architecture.}
\label{fig:NVAE_upd}
\end{center}
\end{figure}

Initially, in the NVAE model, to improve the quality of generation and capture deep dependencies, a vector $h$ was implemented, which was fed to the input at the top of the hierarchy during generation and was a trainable parameter. In our work, for conditional generation in NVAE, we enriched the vector $h$ with the class embedding vector $c$, a trainable parameter, and a vector representation of a specific sample consisting of classes of this sample. In the case of a multi-label problem statement, embedding vector $c$ is the sum of its classes, fed to the input of the encoder and decoder of the cNVAE-ECG model. This representation allows the model, whose layers have been trained to represent the ECG signal space as a whole, to specify the pathology and learn to describe its representation, as well as others. The resulting architecture is shown in Fig.~\ref{fig:NVAE_upd}.

\begin{table*}[!t]
\caption{Selected classes distribution in the PTB-XL training set.}
\centering
\begin{tabular}{|c||c||c||c||c||c||c||c||c||c|}
\hline
Class name                                                & SR                      & 
 MI & LAD                      & TAb                      & LVH                     & AF                      & STach                   & SB                      & IAVB                    \\ \hline
Number of real samples                          & 12,243         & 3,300            & 3,276                     & 1,556                     & 908                    & 849                    & 502                     & 456                     & 445                     \\
Fraction of dataset size,\% & 69 & 18.6 & 18.5 & 8.8 & 5.1 & 4.8 & 2.8 & 2.6 & 2.5 \\
Number of generated samples added to pretrain                      & 0                        & 8,943 & 8,967                    & 10,687                    & 11,335                   & 11,394                   & 11,741                   & 11,787                   & 11,798                   \\ \hline
\end{tabular}
\label{tab:ptb-xl}
\end{table*}

\section{Experimental Setup} \label{pipeline}

The proposed experimental setup consists of several stages. Specifically, we first determined two downstream tasks. The first task is a binary classification of ECG signals to identify pathologies. This approach allows the direct evaluation of the quality of the generated ECG signals by comparing the change in metrics on the test set when adding such signals to the training set. The second task is multi-label classification using transfer learning. Transfer learning is a machine learning technique in which a model previously trained on one dataset improves learning on another dataset~\cite{technologies11020040}. Such an approach, particularly in the classifying ECG signals task, shows a noticeable quality improvement~\cite{transfer_ecg}. It makes it possible to assess a generative model's ability to capture deep signal patterns using them as a pre-train dataset. 

Then, we identified a list of approaches to enrich the training set with generated signals for both tasks. After that, we selected and preprocessed datasets for each of the tasks. Next, we chose a classification model and conducted the testing process, training the model according to the~\cite{DBLP:journals/titb/StrodthoffWSS21}. We describe each step of the pipeline in more detail below.

\subsection{Estimating the quality of generated signals}
We examined several approaches to determine whether adding generated signals directly to the training process would improve the model's final metric on the test set. The validation and test samples were fixed without the addition of any generated data.

\subsubsection{Enrichment of the training dataset for the binary classification task} 
For the first downstream task, we supposed that adding generated data of both classes directly to the training dataset with a similar distribution but with different proportions $n=0.1,0.2,\cdots, 1.0, 1.5, 2.0$ can improve metrics on a test dataset. Also, we tested the addition of samples only of the pathology class in such a way as to minimize class imbalance with proportions $n=0.1,0.2,\cdots ,1.0$. We tested that these improvements can help to improve test metrics on average.

\subsubsection{Enrichment of the pre-train dataset for the transfer learning}
For the second downstream task, we decided to enrich the pretrain dataset so there would be no imbalance of classes. Then, we trained the multi-label classifier on that enriched pretrain and used it for retraining on the other datasets. In this series of tests, we took each retraining dataset with proportions $n=0.1,0.2,\cdots ,1.0$. Finally, we calculated the AUROC metric on the test samples of these datasets for each proportion, after which we averaged the resulting values.

\subsubsection{Comparison with existing ECG generation methods for the same training time} While the main contribution is the cNVAE-ECG model, we also decided to compare our model with conditional versions of two existing generative models for ECG signals, namely WaveGAN* and Pulse2Pulse, mentioned previously. These models belong to the family of GAN-based architectures. This type of architecture is used in the vast majority of ECG signal generation work, so we decided to test the quality of this approach compared to ours. Also, WaveGAN* and Pulse2Pulse are open-source models that can be reproduced. We made these models conditional by adding pathology embedding to the input noise of the generator, similar to cNVAE-ECG.

The Pulse2Pulse model is based on the U-Net architecture by \cite{ronneberger2015unet}, employing 1D-convolutional layers for ECG signal generation. Operating on an $8 \times 5000$ noise vector, matching the output ECG dimension, the Pulse2Pulse network uses five down-sampling blocks with a Leaky ReLU activation followed by five up-sampling blocks. Tab.\ref{tab:p2p} presents the architecture and training hyperparameters we used with the Pulse2Pulse implementation. 

 \begin{table}[h]
\centering
\caption{Pulse2Pulse Training Hyperparameters.}
\label{tab:p2p}
\begin{tabular}{|c||c|}
\hline
Hyperparameter                         & Value  \\ \hline
Batch size                             & 32     \\
Discriminator size                     & 50     \\
Generator size                         & 50     \\
Gradient penalty regularization factor & 10     \\
Learning rate                          & 0.0001 \\
Optimizer                              & Adam   \\
Training epochs                        & 2,400   \\ \hline  
\end{tabular}
\end{table}

WaveGAN* is a WaveGAN-based \cite{donahue2019adversarial} model that takes a 1D $100 \times 1$ random noise vector as input from a uniform distribution with a mean of 0 and standard deviation of 1. This vector undergoes five deconvolution blocks, each consisting of an up-sampling layer, a constant padding layer, a 1D-convolution layer, and a ReLU activation function. The goal is to produce a desired output of $5000 \times 8$ samples, making this implementation deeper than the original architecture for synthesizing audio samples. Tab.\ref{tab:wgs} presents the architecture and training hyperparameters we utilized in the WaveGAN* implementation.

 \begin{table}[!t]
\centering
\caption{WaveGAN* Training Hyperparameters.}
\label{tab:wgs}
\begin{tabular}{|c||c|}
\hline
Hyperparameter                         & Value  \\ \hline
Batch size                             & 32     \\
Discriminator size                     & 50     \\
Generator size                         & 50     \\
Gradient penalty regularization factor & 10     \\
Learning rate                          & 0.0001 \\
Optimizer                              & Adam   \\
Training epochs                        & 2,700   \\ \hline     
\end{tabular}
\end{table}

Our goal was to compare the metrics on the test when adding signals from each architecture separately. We trained cNVAE-ECG for 500 epochs on an A100 GPU, which took about 134 hours. Next, we set the same time for training the WaveGAN* and Pulse2Pulse architectures so that by this time, we could compare the quality of the generated signals by all three models. We made this comparison because running deep learning on GPUs is an environmentally destructive process that leaves a large carbon footprint, so getting the best results as early as possible is essential \cite{Wu2021SustainableAE}.

\subsubsection{Evaluating how the quantity of initial data impacts the quality of generated data} As an additional goal, we tested how the amount of data in the training set affects the performance of both the cNVAE-ECG model and its competitors. Limited data, especially for GAN-like architectures, is a significant challenge in machine learning, and a class imbalance in the dataset can compound this issue. We wanted to investigate how the volume of data across different classes influences the quality of these models, particularly in comparison to VAE-like architectures like cNVAE-ECG.

\subsection{Dataset}

One of the main subtasks was collecting and preparing data for work with subsequent training of the model. We used multiple datasets from the PhysioNet/CinC Challenge 2021~\cite{9662687}. The first source is the data from Physikalisch-Technische Bundesanstalt (PTB-XL)~\cite{article}, consisting of $21837$ 12 lead clinical ECG recordings with a duration of 10 seconds and a frequency of 500 Hz each. 

The second dataset is the Georgia database~\cite{georgia} representing records from the Southeastern United States and containing $10344$ samples, with each record between 5 and 10 seconds long with a sampling frequency of 500 Hz. The third and final source is the Ningbo First Hospital (Ningbo) database~\cite{ningbo} containing 34905 ECGs, with each recording being 10 seconds long with a sampling frequency of 500 Hz.

We used classes from the PTB-XL dataset for training and evaluating binary classification and as a pre-train for multi-label classification tasks since it is actively used in publications \cite{st-net,MEHARI2022105114,SEO2022106858,ALCARAZ2023107115} of ECG signals generation. We used the remaining datasets to evaluate the quality of generated signals in the second downstream task. We randomly split each dataset into training, validation, and test sets.

We selected and preprocessed the described data so that each recording had a sampling rate of 500 Hz and was 10 seconds long. We chose the Myocardial Infarction (MI) for the binary classification task since it is one of the largest classes in the PTB-XL dataset and differs from Sinus Rhythm (SR). For the multi-label classification task, to test the ability to work with unbalanced medium-sized and unpopular classes, we chose Left Axis Deviation (LAD), Left Ventricular Hypertrophy (LVH), Atrial Fibrillation (AF), Sinus Tachycardia (STach), First-degree Atrioventricular Block (IAVB), Sinus Bradycardia (SB) and T-Wave Abnormal (TAb, with all conditions mapped according to SNOMED CT terminology~\cite{El-Sappagh2018}. Also, given the inherent variability and potential measurement noise in physiological data, we applied percentile-based filtering (2.5th to 97.5th) to each lead to exclude outlier values likely to be non-physiological or measurement errors while retaining the majority of valid samples. Table~\ref{tab:ptb-xl} provides statistics of these nine classes compared to SR in the training part of PTB-XL for all downstream tasks. Also, we selected available classes from this list among other datasets.

\subsection{Classification Method and Evaluation Metric}

As model architecture, we used an XResNet1d101 model \cite{he2018bag}, a one-dimensional adaptation of a ResNet architecture \cite{he2015deep}. This architecture demonstrated the best results on an ECG classification task, which was shown in \cite{DBLP:journals/titb/StrodthoffWSS21}. XResNet1d, or 1D-implementation of XResNet, stands as a notable evolution of the conventional ResNet architecture, incorporating distinct features or tweaks contributing to enhanced model performance. These tweaks include modifying the downsampling block of ResNet, adding extra average pooling layers to prevent ignoring input feature maps, and reducing the computational cost of convolutions by changing the size and order of kernels. Tab.\ref{tab:xrs} displays the architecture and training hyperparameters used to implement XResNet1d101. The architecture consists of four blocks with 3, 4, 23, and 3 layers. The training utilized the weighted AdamW optimizer with a learning rate and weight decay set at $1 \times 10^{-3}$ for 50 epochs, and a batch size of 128 was employed.

 \begin{table}[h]
\centering
\caption{XResNet1d101 Training Hyperparameters.}
\label{tab:xrs}
\begin{tabular}{|c||c|}
\hline
Hyperparameter       & Value       \\ \hline
Batch size           & 128         \\
Training epochs      & 50          \\
Block of layers      & 4           \\
Layers in each block & 3, 4, 23, 3 \\
Learning rate        & 0.001       \\
Optimizer            & AdamW       \\
Weight decay         & 0.01      \\ \hline
\end{tabular}
\end{table}

Also, we closely followed the training and evaluation methodology outlined in \cite{DBLP:journals/titb/StrodthoffWSS21}, where AUROC was used to estimate the quality of the model on the test set. Tab.\ref{tab:nvaeecg} presents the architecture and training hyperparameters utilized in the cNVAE-ECG implementation. The hierarchy has three levels in total; the size from the highest to the lowest levels were 5, 10, and 20, respectively. The number of channels in the decoder and encoder is 12, and the number of pre-and post-processing cells, as well as the number of conditional blocks of the encoder and decoder, was 4. The algorithm was trained for 500 epochs with a batch size of 64, with the Adamax optimizer and learning rate equal $1 \times 1^{-3}$.

 \begin{table}[h]
\centering
\caption{cNVAE-ECG Training Hyperparameters.}
\label{tab:nvaeecg}
\begin{tabular}{|c||c|}
\hline
Hyperparameter                                                  & Value     \\ \hline
Training epochs                                                 & 500       \\
Batch size                                                      & 32        \\
Normalizing flows                                               & 0         \\
Latent variable scales                                          & 3         \\
Groups in each scale                                            & 5, 10, 20 \\
Number of channels in encoder and decoder                       & 12        \\
Number of preprocess and postprocess cells                      & 4         \\
Number of cells in conditional encoder and decoder      & 4         \\
Learning rate                                                   & 0.001     \\
Optimizer                                                       & Adamax    \\
Weight decay                                                    & 0.01      \\
Warmup period in epochs & 5         \\ \hline
\end{tabular}
\end{table}

\section{Experiment Results} \label{results}

\begin{table*}[!t]
\centering
\caption{AUROC values (\%) on Georgia dataset depending on pretrain strategy, averaged by proportions.}
\begin{tabular}{|c||c||c||c||c||c|}
\hline
Class name & No pretrain & Original data & Proposed (cNVAE-ECG)& WaveGAN*& Pulse2Pulse \\ \hline
LAD        & 93.80       & 94.67          & \textbf{95.23} & 95.19             & 93.98                \\
TAb        & 89.77       & 92.07          & \textbf{92.40} & 91.23             & 88.40                \\
LVH        & 92.51       & 97.33          & \textbf{97.99} & 96.30             & 94.45                \\
AF         & 91.26       & 93.15          & \textbf{93.67} & 91.44             & 90.71                \\
STach      & 98.46       & \textbf{99.43} & 99.39          & 98.99             & 98.37                \\
SB         & 86.70       & \textbf{88.19} & 87.99          & 86.42             & 83.97                \\
IAVB       & 91.18       & \textbf{93.81} & 93.17          & 92.26             & 89.73       \\ \hline      
\end{tabular}
\label{tab:georgia}
\end{table*}

\begin{table*}[!t]
\centering
\caption{AUROC values (\%) on Ningbo dataset depending on pretrain strategy, averaged by proportions.}
\begin{tabular}{|c||c||c||c||c||c|}
\hline
Class name & No pretrain & Original data       & Proposed (cNVAE-ECG)  & WaveGAN*    & Pulse2Pulse \\ \hline
LAD        & 97.79       & 97.84          & \textbf{98.02} & 97.73             & 97.62                \\
TAb        & 88.86       & 89.55          & \textbf{89.65} & 89.59             & 88.55                \\
LVH        & 91.58       & 91.63          & \textbf{92.08} & 90.73             & 89.67                \\
STach      & 98.49       & 99.02          & \textbf{99.50} & 99.07             & 99.05                \\
SB         & 99.71       & \textbf{99.80} & 99.79          & 99.74             & 99.71                \\
IAVB       & 96.74       & \textbf{97.56} & 97.38          & 97.24             & 96.38         \\ \hline  
\end{tabular}
\label{tab:ningbo}
\end{table*}

\subsection{Quantitative Results}
\subsubsection{Binary classification metrics with proportionally increased entire training dataset} 
\begin{figure}[ht]
\vskip 0.2in
\begin{center}
\centerline{\includegraphics[width=\columnwidth]{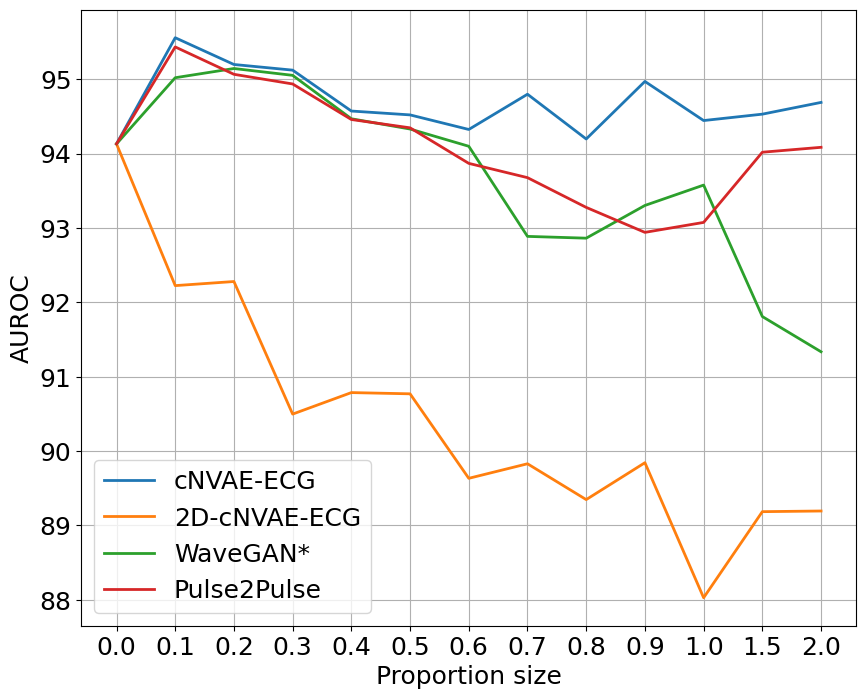}}
\caption{Values of AUROC on the PTB-XL test set for each proportion using four different enrichment methods of both classes in the training dataset.}
\label{fig:hypo1}
\end{center}
\vskip -0.2in
\end{figure}

Fig.~\ref{fig:hypo1} shows that adding data generated by the one-dimensional version of cNVAE-ECG improves the quality of the metric on the test set, in contrast to the 2D-cNVAE-ECG, which degrades the quality of the AUROC. Moreover, in all cases, the result by one-dimensional cNVAE-ECG is better than using WaveGAN* and Pulse2Pulse models. The best metrics are achieved by adding generated data with a proportion of 0.1, 0.2, 0.3, and 0.9, with the most increase of $1.5\%$ compared to the AUROC value without any data addition. This result can be explained by the fact that adding a small proportion of generated data introduces regularization, making the model more robust when working with ECG signals, and large proportions, due to the more significant amount of data, generally improve the model's ability to capture patterns in the signals.

\subsubsection{Binary classification metrics with proportionally increased Myocardial Infarction class} 
\begin{figure}[ht]
\vskip 0.2in
\begin{center}
\centerline{\includegraphics[width=\columnwidth]{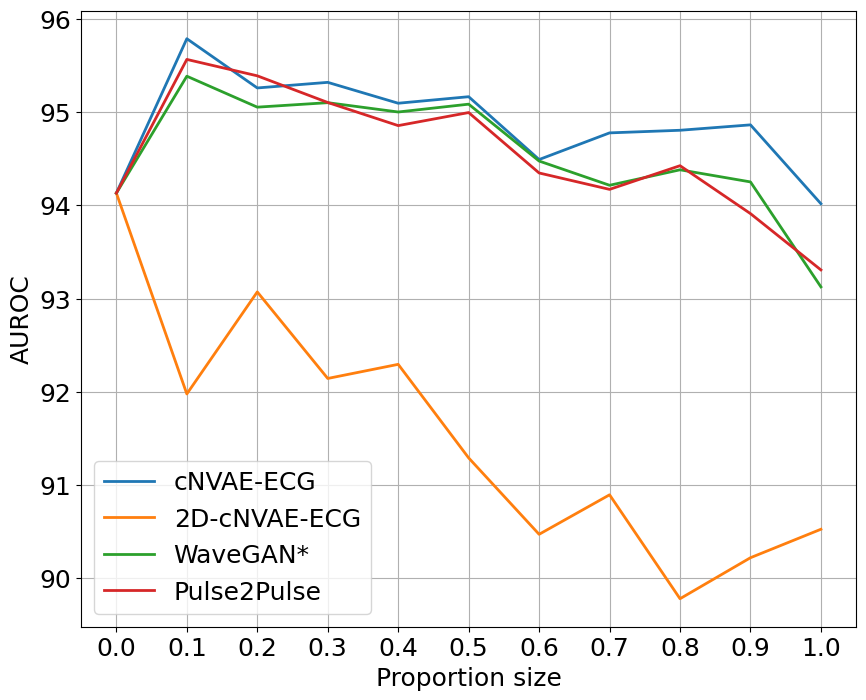}}
\caption{Values of AUROC on the PTB-XL test set for each proportion using four different enrichment methods of the Myocardial Infarction class in the training dataset.}
\label{fig:hypo2}
\end{center}
\vskip -0.2in
\end{figure}

Only generated signals of the Myocardial Infarction class with a given proportion to lower imbalance were added to the training set. The results are shown in the Fig.~\ref{fig:hypo2}. Increasing only the Myocardial infarction in the training set demonstrates the best AUROC with small proportions, with the most significant increase of $1.8\%$ compared to the AUROC value without any data addition. Then, the metric value gradually decreases but still shows better results than without generated data addition. Generally, augmenting a smaller class and reducing the original imbalance with new objects improves the classification metric, except for the 2D-cNVAE-ECG. However, as the proportions of the Myocardial infarction increase, the AUROC decreases. This decrease may mean that the models could only partially capture the distribution of the Myocardial infarction due to class imbalance with the limited number of unique signals of this pathology represented in the dataset.

\subsubsection{Multi-label classification metrics for each proportion of Georgia training dataset} 
Table~\ref{tab:georgia} shows that using a pretrain for the Georgia dataset generally improves the model quality for all classes. Particularly, pretrain enriched with cNVAE-ECG shows the best results among GAN-like methods, surpassing basic pretrain for the four most popular (according to Table~\ref{tab:ptb-xl}) classes in the cNVAE-ECG training set. This observation suggests that cNVAE-ECG captured the dependencies for rare classes less strongly than for the more popular ones. Additionally, in some cases Pulse2Pulse and WaveGAN* models produces worse results than without using pre-training at all.

\subsubsection{Multi-label classification metrics for each proportion of Ningbo training dataset} 
Table~\ref{tab:ningbo} shows that for most classes, adding results generated by cNVAE-ECG and GAN-like models gives an acceptable increase in quality on the Ningbo test set compared to metrics without pretrain. The cNVAE-ECG model again shows the best result on all classes compared to the other generative methods. However, it results worse in rare classes, such as SB and IAVB, than with the original pretrain. Considering the previous results, we can conclude that cNVAE-ECG does not have enough data from these classes to generate ECG signals of these types successfully.

\begin{figure}[!t]
\centering
\subfloat[]{\includegraphics[width=1.7in]{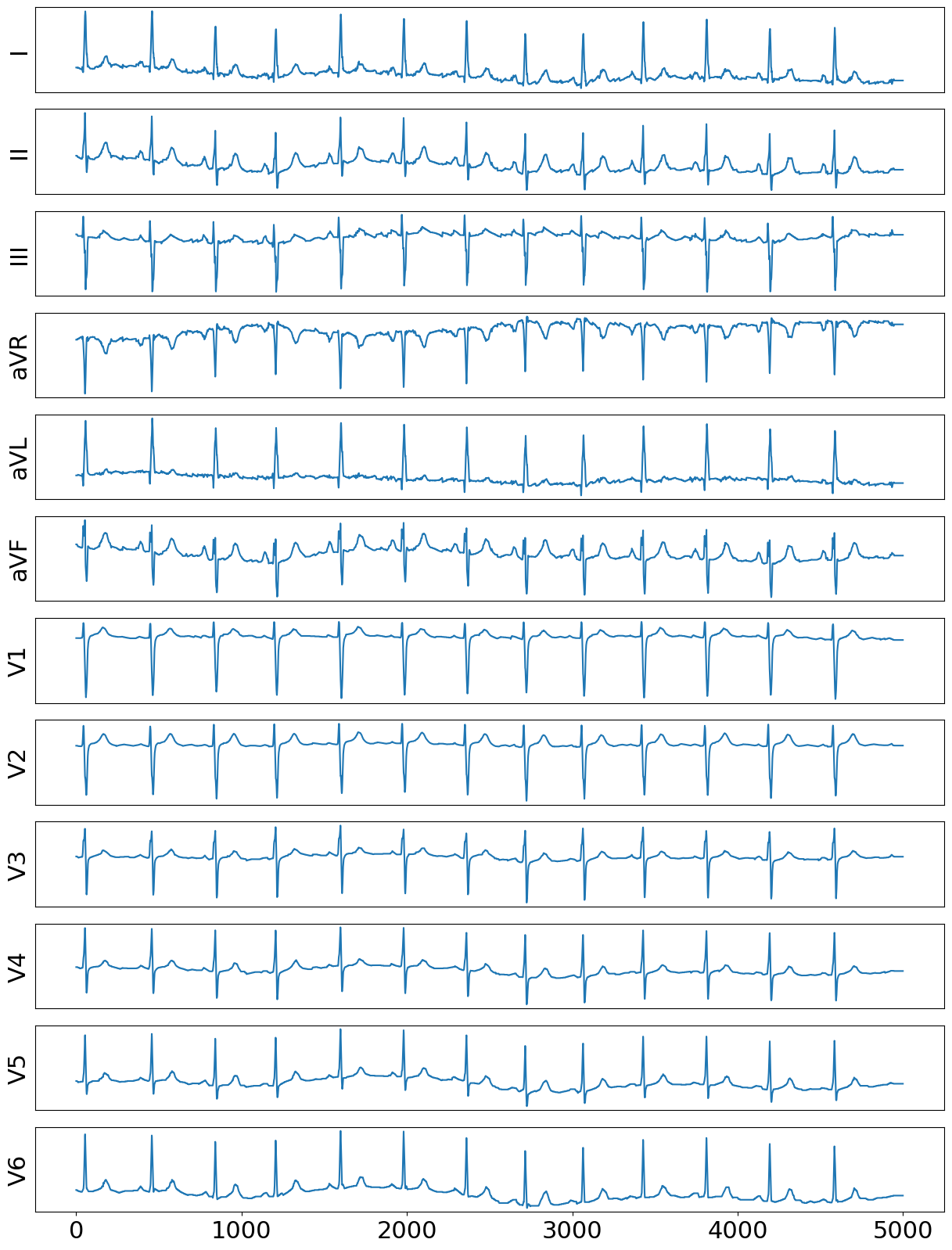}%
\label{fig_first_case}}
\hfil
\subfloat[]{\includegraphics[width=1.7in]{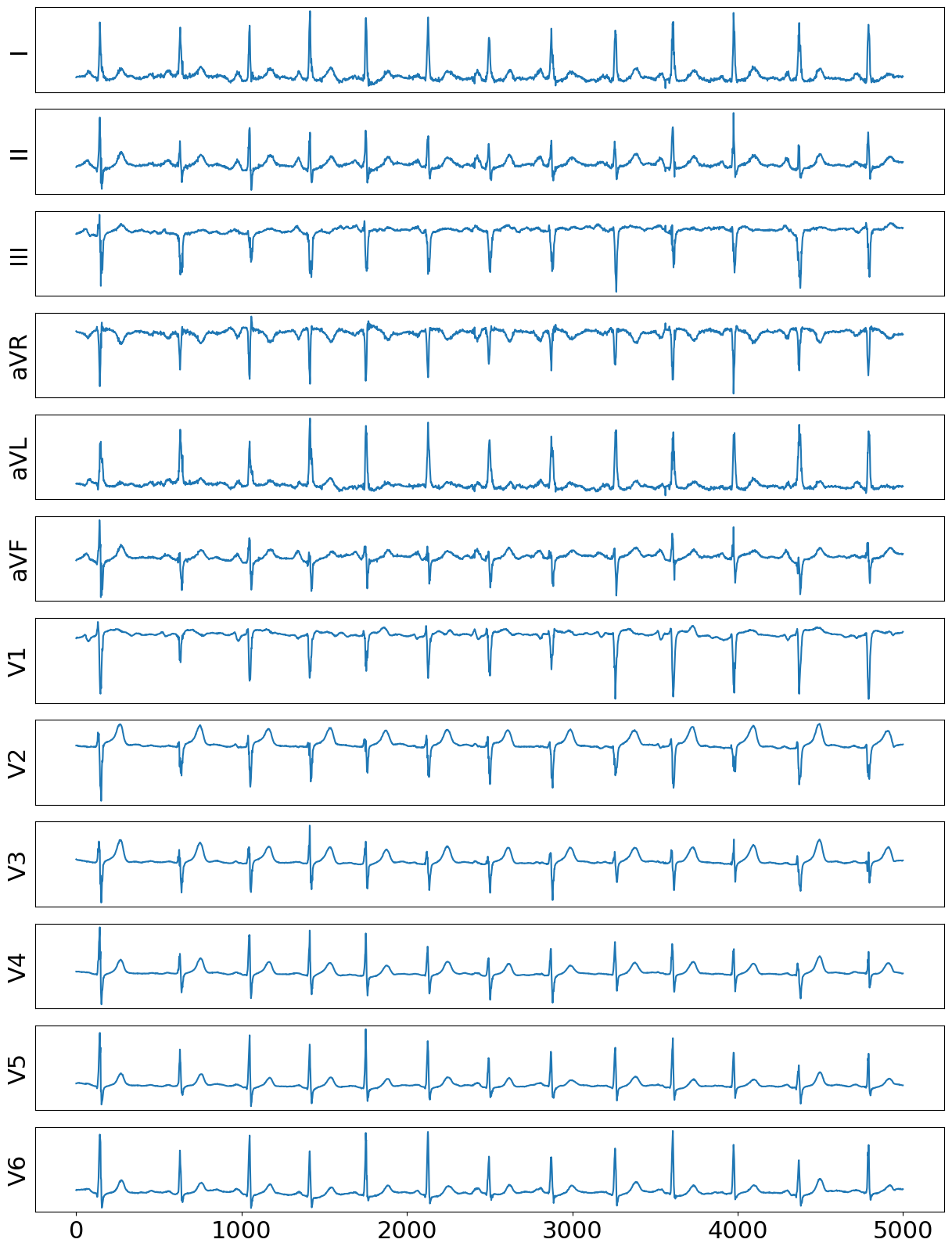}%
}
\caption{Real (a) and generated by cNVAE-ECG (b) ECG signal for the Sinus rhythm class.}
\label{fig:0_good}
\end{figure}

\begin{figure}[!t]
\centering
\subfloat[]{\includegraphics[width=1.7in]{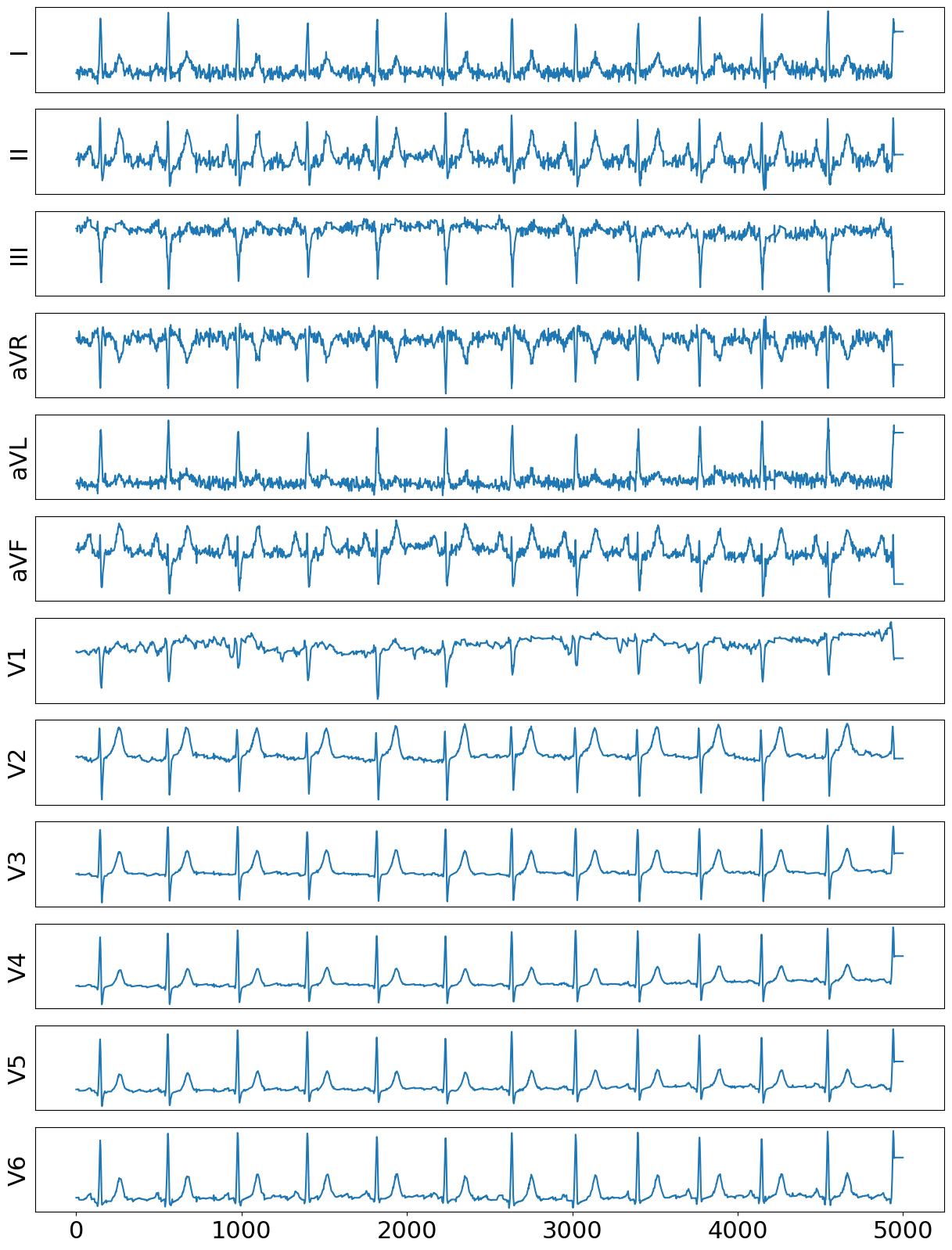}%
\label{fig_first_case}}
\hfil
\subfloat[]{\includegraphics[width=1.7in]{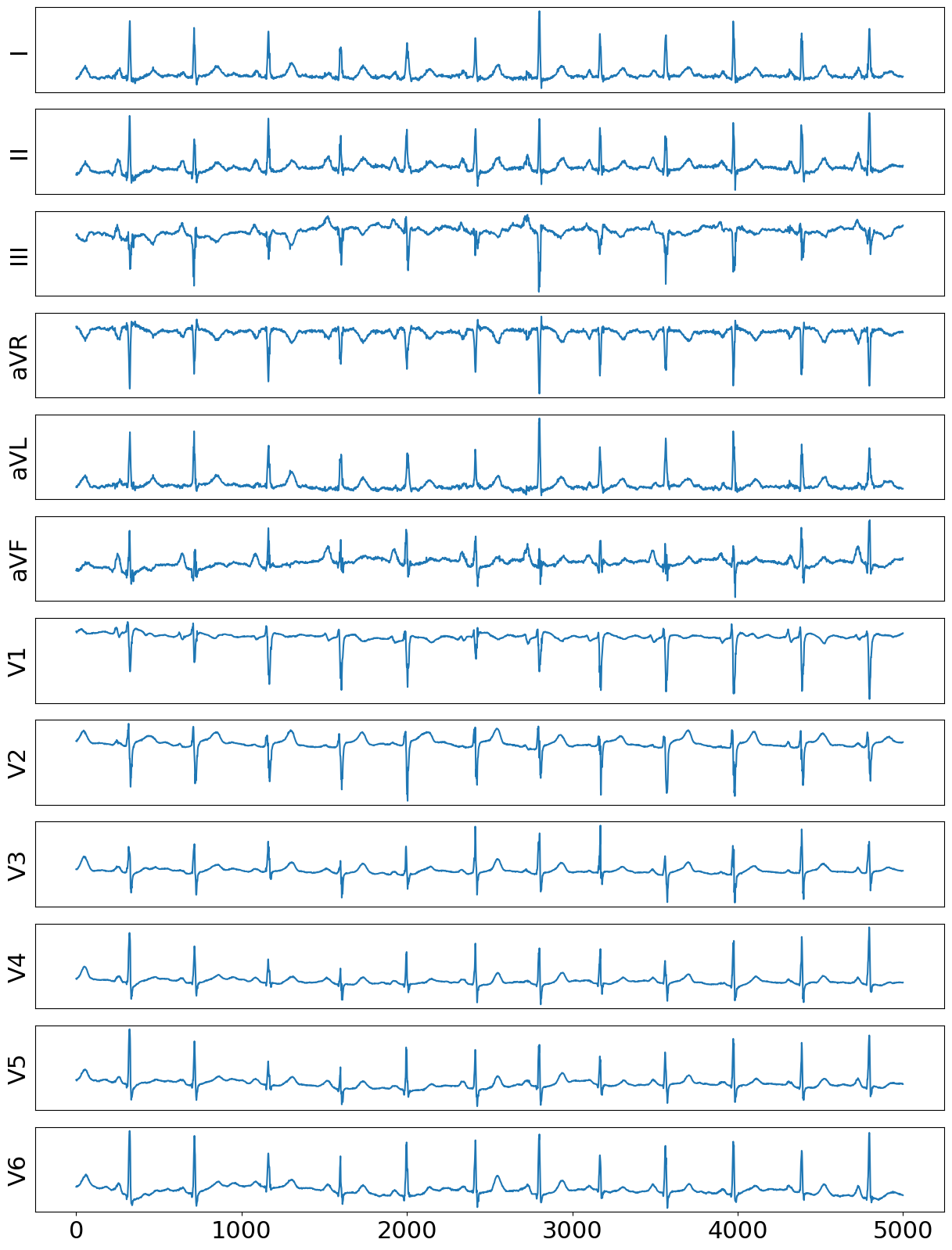}%
}
\caption{Real (a) and generated by cNVAE-ECG (b) ECG signal for Myocardial infarction.}
\label{fig:1_good}
\end{figure}

\subsection{Qualitative Results}

\begin{figure}[ht]
\vskip 0.2in
\begin{center}
\centerline{\includegraphics[width=\columnwidth]{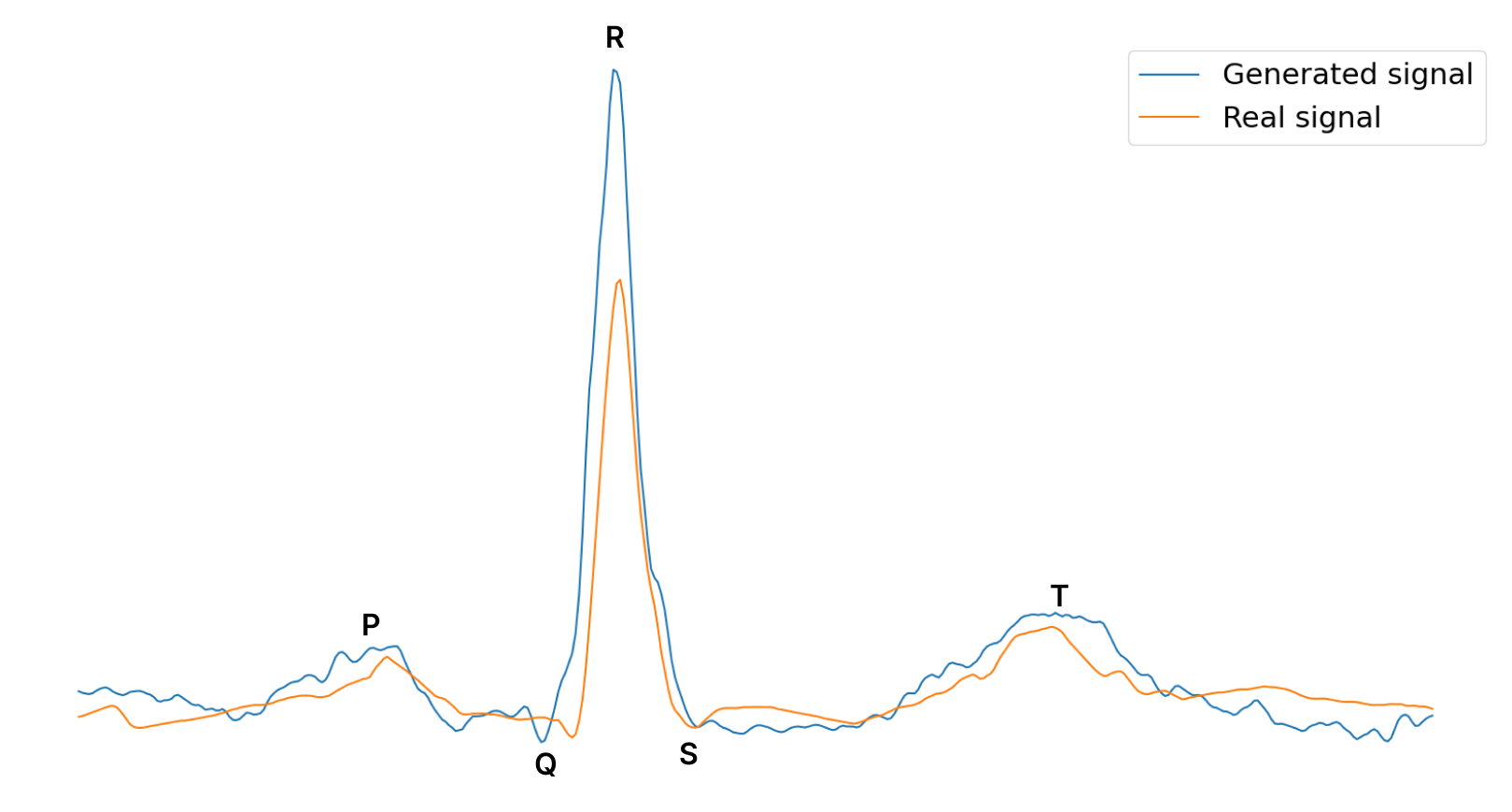}}
\caption{Comparison between real and generated one heartbeat of Lead I for the Sinus rhythm class.}
\label{fig:one_beat}
\end{center}
\vskip -0.2in
\end{figure}

Although the main goal was to improve classification metrics on downstream tasks, obtaining data similar to ECG signals and reflecting the training set distribution and diversity was also essential. Fig.~\ref{fig:0_good} and Fig.~\ref{fig:1_good} show the example of generated ECG signals for Sinus rhythm and Myocardial infarction classes, respectively, and the closest actual signal from the training set. Although the generated signals by cNVAE-ECG have some artifacts, they display the distribution of both classes, thereby showing that the cNVAE-ECG model can generate plausible ECG signals. Also, we provided a more detailed comparison of one heartbeat for the lead I of Sinus rhythm ECG in Fig.~\ref{fig:one_beat}. It suggests that the cNVAE-ECG successfully reproduces the core structure and characteristics of the real ECG signal, such as P, Q, R, S, and T peaks~\cite{berkaya2018survey}.

\section{Conclusions}\label{conclusion}
We developed a cNVAE-ECG model to generate ECG signals with 12 leads and a standard 10-second duration based on NVAE architecture. We also tested the quality of generated signals on two downstream tasks: binary pathology classification task and multi-label classification using transfer learning. We compared the results with the existing methods for generating ECG signals based on GANs, namely WaveGAN* and Pulse2Pulse, and a two-dimensional version of cNVAE-ECG. For the binary classification task, we checked the improvement of test metrics when adding generated data to the training set: proportional enrichment of the entire training dataset and enrichment of only classes with few examples. 

Adding both generated classes proportionally to the original data distribution improved the test metric, showing the best result compared to no data addition and among GAN-like models, as did adding only positive examples to eliminate class imbalance. These results show that it is essential to enrich not only the classes with a small number of examples but also the larger classes so that the model can more accurately and deeply reflect the dependencies in the data. It is also essential to consider the data proportions, focusing on small proportions to add only positive classes and considerable proportions to enrich both classes.

Also, we pretrained a classifier on the PTB-XL dataset for seven medium-sized and unpopular classes for the multi-label classification task. After that, we retrained the obtained model in different variations: without pre-train, using original pretrain, and with pretrain augmented with data generated by used generative models. We selected the Georgia and Ningbo datasets for retraining, calculated the quality, and compared the mean results by the proportions of these datasets. Metrics demonstrated that adding generated data by the cNVAE-ECG model gives a better gain in metrics, especially among GAN-like architectures, than without pretrain. However, the metric values obtained when solving downstream tasks are lower for rare classes in the training dataset than those without adding the generated signals. These results suggest the potential for developing VAE-based generative methods for underrepresented classes.

Despite the noise and artifacts, the generated data is similar to ECG signals, preserving their properties and diversity. Considering the described results, we plan to have these data validated by medical professionals to ensure its accuracy and applicability. Following validation, we plan to conduct pilot projects using the generated data to test the algorithms in real-world scenarios, incorporating federated learning techniques~\cite{9344971}. These projects aim to advance the field in automatic ECG analysis, developing approaches that show promising results on actual data and revile improved generalization capabilities.

\bibliographystyle{IEEEtran}
\bibliography{main}

\end{document}